\documentclass[twocolumn,twoside,aps,pra,superscriptaddress,floatfix]{revtex4-1}
\usepackage{epsf,epsfig,amsfonts,amsmath,amssymb,graphicx,times}
\usepackage[utf8]{inputenc}
\usepackage[OT1]{fontenc}
\usepackage{mathtools,color,stmaryrd,hyperref}
\begin{document}
\title{Generalized security analysis framework for continuous-variable quantum key distribution}
\author{Vladyslav C. Usenko}
\affiliation{Department of Optics, Palacký University, 17. listopadu 50, 772~07 Olomouc, Czech Republic}

\begin{abstract}
Security of practical continuous-variable quantum key distribution is addressed and a security analysis framework, which does not rely on phase-space symmetries of signal states and correlations, is developed. In a general purification-based approach, following optimality of Gaussian collective attacks, it is suggested to find an equivalent generally mixed two-mode state shared between the trusted parties and then purifying it using Bloch-Messiah decomposition. This allows to assess security of the schemes with arbitrary parameters, which can be typically expected in experiments. It also allows to theoretically predict the role of asymmetries of signals and correlations on security of the protocols. The method can be used for security analysis of practical continuous-variable schemes directly from the measured data without any symmetrization assumptions.
\end{abstract}
\maketitle

\section{Introduction}
Continuous-variable (CV) quantum key distribution (QKD) (see \cite{Pirandola2019} for review) is aimed at efficient realization of quantum cryptography using off-the-shelf optical communication components. Security of CV QKD was established against individual, collective and general attacks in an asymptotic as well as finite-size regime for some protocols \cite{Diamanti2015}. Realization of CV QKD is typically performed using prepare-and-measure schemes, when an optimized Gaussian quadrature modulation is applied to coherent or squeezed signal states \cite{Cerf2001,Grosshans2002}. The core of security analysis of Gaussian CV QKD is evaluation of Holevo bound, which upper limits eavesdropper’s information, and follows optimality of collective Gaussian attacks \cite{Navascues2006,Garcia2006}. In a general case of lossy and noisy quantum channel an eavesdropper Eve is assumed to be able to purify all the noise related to the channel. This allows evaluating the Holevo bound from the entropies of the state, which is shared between the trusted parties, namely Alice and Bob. In order to distinguish between trusted (related to Alice's and Bob's device imperfections such as noise of the signal, modulator and detectors) and untrusted noise (related to the untrusted quantum channel), the state shared between Alice and Bob must be theoretically purified \cite{Usenko2016}. Then all the additional impurities can be attributed to the channel and contribute to Eve's information. State preparation is typically purified using an equivalent two-mode pure entangled state \cite{Grosshans2003a}. However, such approach supposes symmetries of modulated states and correlations in the phase space and, in case of squeezed states, fixes the modulation variance to the level of signal squeezing (the latter limitation can be removed by generalized pure-state preparation schemes \cite{Usenko2011,Derkach2017}, which are still limited by the two-mode state purity). Moreover, trusted noise in this case should be still added to the equivalent pure state using coupling to another pure entangled two-mode state \cite{Lodewyck2007}, which can be also approximately lossless \cite{Usenko2010a}. In order to overcome the limitations and avoid approximations, we develop a generalized approach to security analysis of CV QKD, which is based on construction of an equivalent generally mixed two-mode state from the actually measured variances and correlations in each of the quadratures, and then purification of the obtained state applying Bloch-Messiah reduction to CV states \cite{Braunstein2005a}. This allows to assess security of practical CV QKD schemes and study the role of asymmetries.

\section{Generalized purification scheme}
\begin{figure}
\centering
\vspace{10px}
\includegraphics[width=0.45\textwidth]{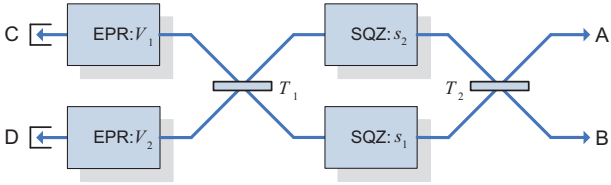}	
\caption{Four-mode purification scheme for an arbitrary two-mode state in modes $A,B$ using two two-mode squeezed vacuum states (EPR) with variances $V_{1,2}$, two single-mode squeezers (SQZ) with squeezing parameters $s_{1,2}$ and two variable couplers before and after the squeezers with transmittance values $T_{1,2}$. Auxiliary modes $C,D$ are kept in the purification, but are not accessible neither two the trusted parties, nor to Eve. State of the modes $A,B$ is exactly as in the two-mode generally mixed state (\ref{actmat}), equivalent to imperfect prepare-and-measure scheme.
\label{gen_prep}}
\end{figure}

After the preparation and measurement, trusted parties in CV QKD perform channel estimation (in the finite-size regime of limited data ensembles it leads to confidence intervals on the estimated parameters depending on the acceptable probability of failure \cite{Leverrier2010} and can be optimized \cite{Ruppert2014}, which we however omit here, focusing on the asymptotic security, that can then be directly extended to finite-size regime using existing techniques). Knowing channel transmittance and excess noise, the trusted parties can reconstruct the state (quadrature variances) prior to the channel as well as correlations between the data. We assume that the reconstructed variances of the modulated signal entering the channel are $V_B^{(x,p)}$ in each of the complementary quadratures (namely $x=a^{\dag}+a,p=i(a^{\dag}-a)$, expressed in terms of the quantum operators of a given mode), Alice possesses modulation data with variance $V_M^{(x,p)}$ and the respective correlations generally are $C_{MB}^{(x,p)}$ (note that typically $C_{MB}^{(x,p)}=V_M^{(x,p)}$, but a more general situation can be foreseen, also state monitoring after the modulation by a tap-off or optical switching may be needed). We build an equivalent two-mode state of the form
\begin{equation}
\label{genmat}
\gamma_{AB}=\left(
\begin{array}{cccc}
 V_A^{(x)} & 0 & C_{AB}^{(x)} & 0 \\
 0 & V_A^{(p)} & 0 & C_{AB}^{(p)} \\
 C_{AB}^{(x)} & 0 & V_B^{(x)} & 0 \\
 0 & C_{AB}^{(p)} & 0 & V_B^{(p)} \\
\end{array}
\right),
\end{equation} 
which should i) contain the same mode ($B$), described by $V_B^{(x,p)}$, as was propagating through the channel, so that the entanglement-based and the prepare-and-measure preparation schemes are indistinguishable for Eve, who observes the same reduced state in the channel, and ii) result in the same signal states conditioned by Alice's measurement on her mode. For the coherent-state protocol, which is equivalent to heterodyne measurement of Alice's mode of the two-mode state (\ref{genmat}), the latter condition leads to the simple ratio
\begin{equation}
\frac{C_M^2}{V_M}=\frac{C_{AB}^2}{V_A+1},
\end{equation}
where we omitted quadrature upper-scripts for simplicity (i.e., the expression is the same in both quadratures), and which leads to equivalence of the mutual information in prepare-and-measure and entanglement-based schemes. However, this keeps one of the parameters undefined and we therefore symmetrize the covariance matrix by putting $V_A=V_B$ in both x and p, hence arriving at the matrix
\begin{widetext}
\begin{equation}
\label{actmat}
\gamma_{AB}=\left(
\begin{array}{cccc}
 V_B^{(x)} & 0 & C_{MB}^{(x)}\sqrt{\frac{1+V_B^{(x)}}{V_M^{(x)}}} & 0 \\
 0 & V_B^{(p)} & 0 & C_{MB}^{(p)}\sqrt{\frac{1+V_B^{(p)}}{V_M^{(p)}}} \\
 C_{MB}^{(x)}\sqrt{\frac{1+V_B^{(x)}}{V_M^{(x)}}} & 0 & V_B^{(x)} & 0 \\
 0 & C_{MB}^{(p)}\sqrt{\frac{1+V_B^{(p)}}{V_M^{(p)}}} & 0 & V_B^{(p)} \\
\end{array}
\right),
\end{equation} 
\end{widetext}
which satisfies the equivalence conditions and represents the general state preparation prior to the channel. In the case, when $V_B^{(x)}=V_B^{(p)} \equiv V$ and $C_{MB}^{(x)}=C_{MB}^{(p)}=V_M^{(x)}=V_M^{(p)}=V-1$, state preparation is perfect and (\ref{actmat}) turns into one for the standard pure two-mode squeezed vacuum state. However, it is not necessarily the case and generally for a arbitrary set of parameters the state (\ref{actmat}) is not pure and must be purified to account for the trusted noise. This can be done by applying Bloch-Messiah decomposition to CV states \cite{Braunstein2005a} using an optical network as given in Fig. \ref{gen_prep}, which consists of two two-mode entangled sources, two variable couplers and two squeezers set so that modes $A,B$ of the four modes $A,B,C,D$ will be produced exactly in the state given by (\ref{actmat}). The modes $C,D$ are kept in the purification, but are not accessible neither two the trusted parties, nor to Eve. This approach was previously used in entanglement-based CV QKD \cite{Madsen2012} to find purification of generally noisy two-mode entangled states.
In the case of the state (\ref{actmat}), which consists of four unknown parameters, the scheme in Fig. \ref{gen_prep} can be 
simplified by setting $T_1=1,T_2=1/2$. The parameters of the two-mode state (\ref{actmat}) are then related to the purification scheme through the system of four equations
\begin{equation}
\left.\begin{aligned}
V_B^{(x,p)}=\frac{1}{2}\big(e^{\pm s_2}V_2+e^{\pm s_1}V_1\big) \\
C_{AB}^{(x,p)}=\frac{1}{2}\big(e^{\pm s_2}V_2-e^{\pm s_1}V_1\big).
\end{aligned}\right.
\end{equation}
The system can be solved analytically to find the parameters of purification scheme in Fig. \ref{gen_prep}, and the physically valid solution reads
\begin{equation}
\left.\begin{aligned}
V_1 &=\sqrt{\Big(V_B^{(x)}-C_{AB}^{(x)}\Big)\Big(V_B^{(p)}-C_{AB}^{(p)}\Big)} \\
V_2 &=\sqrt{\Big(V_B^{(x)}+C_{AB}^{(x)}\Big)\Big(V_B^{(p)}+C_{AB}^{(p)}\Big)} \\
s_1 &=\frac{1}{4}\log{\frac{V_B^{(p)}-C_{AB}^{(p)}}{V_B^{(x)}-C_{AB}^{(x)}}} \\
s_2 &=\frac{1}{4}\log{\frac{V_B^{(p)}+C_{AB}^{(p)}}{V_B^{(x)}+C_{AB}^{(x)}}}
\end{aligned}\right.
\end{equation}
Knowing these parameters, we can build the explicit four-mode pure-state covariance matrix $\gamma_{ABCD}$, containing $\gamma_{AB}$ (\ref{actmat}) as a sub-matrix. Now mode $A$ is accessible to Alice for a heterodyne measurement, while the mode $B$ travels through a generally lossy and noisy channel (which makes the overall state in modes $ABCD$ mixed), and is measured by Bob. Then, in the purification-based approach, the von Neumann (quantum) entropy $S(E)$ of the state, which is available to Eve for the collective measurement, will be $S(E)=S(ABCD)$, i.e., the same as entropy of the four-mode state shared between Alice and Bob over the quantum channel. Similarly, the quantum entropy of the state conditioned on the Bob's measurement (relevant in the robust reverse reconciliation scenario) is $S(E|B)=S(ACD|B)$ and the Holevo bound can be obtained as $\chi_{BE}=S(ABCD)-S(ACD|B)$, calculated from the covariance matrices of the respective states (see details on Gaussian CV QKD security analysis using covariance matrix formalism in \cite{Usenko2016}). On the other hand, the mutual information between the trusted parties can be obtained directly from the measured data and established correlations as
\begin{equation}
I_{AB}^{(hom)}=\frac{1}{2}\log{\Bigg(1+\frac{C_{MB}^{'2}}{V_B'V_M-C_{MB}^{'2}}\Bigg)}
\end{equation}
in whatever quadrature measured by the homodyne detection on Bob's side or as 
\begin{eqnarray}
I_{AB}^{(het)}=\frac{1}{2}
\log
	\Bigg[\Bigg(1+\frac{
			C_{MB}^{(x)'2}
			}{
			\Big(V_B^{(x)'}+1\Big)V_M^{(x)}-C_{MB}^{(x)'2}
			}\Bigg)\times{}
\nonumber\\
{}\times\Bigg(1+\frac{
			C_{MB}^{(p)'2}
			}{
			\Big(V_B^{(p)'}+1\Big)V_M^{(p)}-C_{MB}^{(p)'2}
			}\Bigg)\Bigg]&
\end{eqnarray}
for the heterodyne detection, where $V_B^{(x,p)'}$ and $C_{MB}^{(x,p)'}$ are the measured parameters after the channel influence, the bases of the logarithms to be defined by the used encoding (e.g. nats for base $e$ or bits for base $2$). This provides all the ingredients needed to evaluate the lower bound on secure key rate in the case of collective attacks, which in the asymptotic regime reads $K=max\{0,I_{AB}-\chi_{BE}\}$, where $I_{AB}$ is the mutual information between Alice and Bob in either Bob's measurement scenario. The described method allows assessing security of CV QKD with arbitrary and generally asymmetrical signal states, correlations and modulation data, while taking into account trusted imperfect preparation.

\section{Role of asymmetries}

\begin{figure}[h]
\centering
\includegraphics[width=0.45\textwidth]{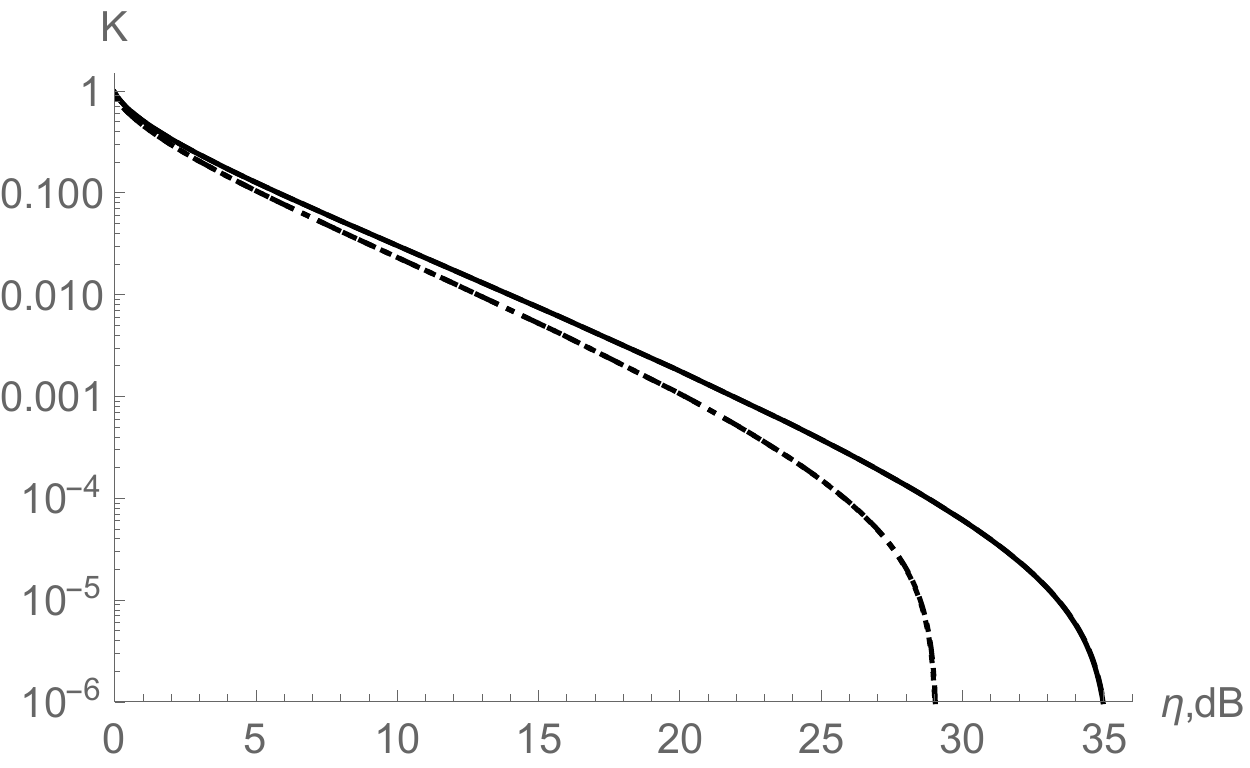} \\
\includegraphics[width=0.45\textwidth]{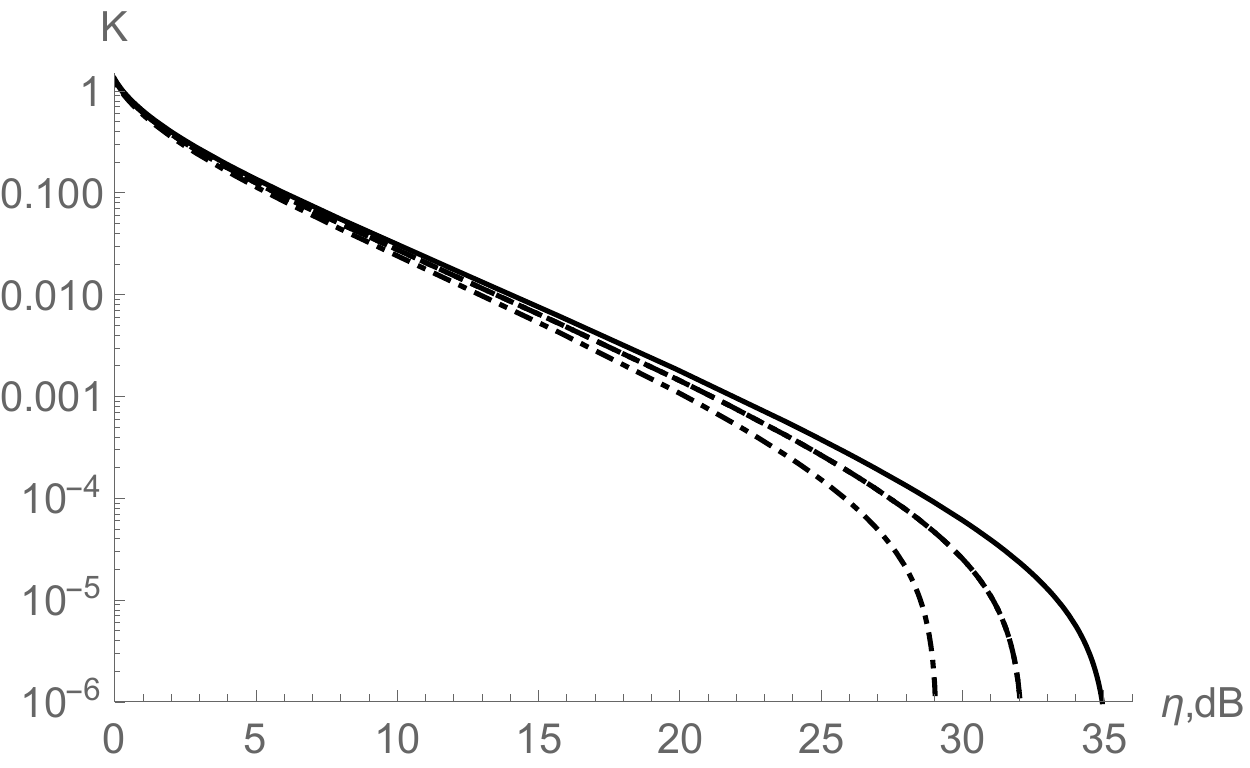}	
\caption{Lower bound on secure key rate for coherent-state Gaussian CV QKD versus channel attenuation $\eta$ in dB scale for homodyne (upper panel) and heterodyne (lower panel) detection on the Bob's side. Modulation variance $V_M^{(x,p)}=9$, correlation $C_{MB}^{(x,p)}=\pm 9$, Bob's state variance prior to channel $V_B^{(x,p)}=10$ (perfect realization, solid lines), $V_B^{(x)}=10,V_B^{(p)}=10.1$ (phase-sensitive preparation noise 0.1 SNU in p-quadrature, dashed lines), $V_B^{(x)}=10.1,V_B^{(p)}=10$ (phase-sensitive  preparation noise 0.1 SNU in x-quadrature, dotted lines), $V_B^{(x,p)}=10.1$ (phase-insensitive  preparation noise 0.1 SNU in both quadratures, dot-dashed lines), channel noise related to the channel input $7\%$ SNU. Homodyne protocol is sensitive only to  preparation noise in the measured quadrature (solid and dashed lines and dotted and dot-dashed lines overlap pairwise), heterodyne protocol is similarly degraded by phase-sensitive  preparation noise (dotted and dashed lines overlap) and is further undermined by phase-insensitive preparation noise.
\label{plot_noise}}
\end{figure}

To illustrate applicability of the generalized security framework, we use the above described methodology to study the role of asymmetries in the coherent-state protocol, which can be caused by generally phase-sensitive trusted preparation noise (previously only phase-insensitive preparation excess noise was studied in CV QKD \cite{Usenko2010a,Usenko2016}). We address coherent-state protocol with homodyne as well as with heterodyne detection (i.e., the non-switching protocol \cite{Weedbrook2004}) and reverse reconciliation, realized over lossy and noisy channels, and visualize the effect of symmetrical and asymmetrical state preparation noise in Fig. \ref{plot_noise} (note that we don't take finite-size effects, including imperfect post-processing, into account here in order to clearly demonstrate the role of asymmetrical trusted noise, the additional effects would just scale the dependencies down).
It is evident from the plots, that while phase-sensitive preparation noise in the unmeasured quadrature does not play role in the coherent-state protocol with the homodyne detection, it does limit the key rate (and, respectively, the security distance of the protocol) when appears in the measured one (here we assume, that complementary quadrature is measured only for the purposes of the full channel estimation) - no matter, if the noise is symmetrical, or not. In the case of heterodyne protocol, however, the phase-sensitive asymmetrical preparation noise in either of the quadratures reduces the key rate (or secure distance), while the phase-insensitive symmetrical preparation noise leads to further degradation of the protocol. 

The above given examples show how the asymmetries in the CV QKD protocol realizations should not be overlooked in the practical situations and may also lead to tighter bounds on Eve's information. 

\section{Conclusions}
We have developed the generalized security analysis framework for CV QKD, allowing to apply the most general purification-based security analysis to an arbitrary protocol with signals, modulation and correlations being asymmetrical in the phase space. This allows evaluating security of the practical implementations of CV QKD using imperfect devices. Our methodology can be also used to predict the performance of the protocols in the presence of asymmetries. The result paves the way to practical implementation of CV QKD with imperfect devices.

\acknowledgements
Author acknowledges support from EU H2020 grant No 820466 (CiViQ), project 19-23739S of the Czech Science Foundation and project LTC17086 of INTER-EXCELLENCE program of the Czech Ministry of Education.


\begin{thebibliography}{16}
\expandafter\ifx\csname natexlab\endcsname\relax\def\natexlab#1{#1}\fi
\expandafter\ifx\csname bibnamefont\endcsname\relax
  \def\bibnamefont#1{#1}\fi
\expandafter\ifx\csname bibfnamefont\endcsname\relax
  \def\bibfnamefont#1{#1}\fi
\expandafter\ifx\csname citenamefont\endcsname\relax
  \def\citenamefont#1{#1}\fi
\expandafter\ifx\csname url\endcsname\relax
  \def\url#1{\texttt{#1}}\fi
\expandafter\ifx\csname urlprefix\endcsname\relax\def\urlprefix{URL }\fi
\providecommand{\bibinfo}[2]{#2}
\providecommand{\eprint}[2][]{\url{#2}}

\bibitem[{\citenamefont{Pirandola et~al.}(2019)\citenamefont{Pirandola,
  Andersen, Banchi, Berta, Bunandar, Colbeck, Englund, Gehring, Lupo, Ottaviani
  et~al.}}]{Pirandola2019}
\bibinfo{author}{\bibfnamefont{S.}~\bibnamefont{Pirandola}},
  \bibnamefont{et~al.}, \bibinfo{journal}{arXiv preprint arXiv:1906.01645}
  (\bibinfo{year}{2019}), 

\bibitem[{\citenamefont{Diamanti and Leverrier}(2015)}]{Diamanti2015}
\bibinfo{author}{\bibfnamefont{E.}~\bibnamefont{Diamanti}} \bibnamefont{and}
  \bibinfo{author}{\bibfnamefont{A.}~\bibnamefont{Leverrier}},
  \bibinfo{journal}{Entropy} \textbf{\bibinfo{volume}{17}},
  \bibinfo{pages}{6072} (\bibinfo{year}{2015}).

\bibitem[{\citenamefont{Cerf et~al.}(2001)\citenamefont{Cerf, Levy, and
  Van~Assche}}]{Cerf2001}
\bibinfo{author}{\bibfnamefont{N.~J.} \bibnamefont{Cerf}},
  \bibinfo{author}{\bibfnamefont{M.}~\bibnamefont{Levy}}, \bibnamefont{and}
  \bibinfo{author}{\bibfnamefont{G.}~\bibnamefont{Van~Assche}},
  \bibinfo{journal}{Phys. Rev. A} \textbf{\bibinfo{volume}{63}},
  \bibinfo{pages}{052311} (\bibinfo{year}{2001}).

\bibitem[{\citenamefont{Grosshans and Grangier}(2002)}]{Grosshans2002}
\bibinfo{author}{\bibfnamefont{F.}~\bibnamefont{Grosshans}} \bibnamefont{and}
  \bibinfo{author}{\bibfnamefont{P.}~\bibnamefont{Grangier}},
  \bibinfo{journal}{Phys. Rev. Lett.} \textbf{\bibinfo{volume}{88}},
  \bibinfo{pages}{057902} (\bibinfo{year}{2002}).

\bibitem[{\citenamefont{Navascu{\'e}s et~al.}(2006)\citenamefont{Navascu{\'e}s,
  Grosshans, and Acin}}]{Navascues2006}
\bibinfo{author}{\bibfnamefont{M.}~\bibnamefont{Navascu{\'e}s}},
  \bibinfo{author}{\bibfnamefont{F.}~\bibnamefont{Grosshans}},
  \bibnamefont{and} \bibinfo{author}{\bibfnamefont{A.}~\bibnamefont{Acin}},
  \bibinfo{journal}{Phys. Rev. Lett.} \textbf{\bibinfo{volume}{97}},
  \bibinfo{pages}{190502} (\bibinfo{year}{2006}).

\bibitem[{\citenamefont{Garcia-Patron and Cerf}(2006)}]{Garcia2006}
\bibinfo{author}{\bibfnamefont{R.}~\bibnamefont{Garcia-Patron}}
  \bibnamefont{and} \bibinfo{author}{\bibfnamefont{N.~J.} \bibnamefont{Cerf}},
  \bibinfo{journal}{Phys. Rev. Lett.} \textbf{\bibinfo{volume}{97}},
  \bibinfo{pages}{190503} (\bibinfo{year}{2006}).

\bibitem[{\citenamefont{Usenko and Filip}(2016)}]{Usenko2016}
\bibinfo{author}{\bibfnamefont{V.~C.} \bibnamefont{Usenko}} \bibnamefont{and}
  \bibinfo{author}{\bibfnamefont{R.}~\bibnamefont{Filip}},
  \bibinfo{journal}{Entropy} \textbf{\bibinfo{volume}{18}}, \bibinfo{pages}{20}
  (\bibinfo{year}{2016}).

\bibitem[{\citenamefont{Grosshans et~al.}(2003)\citenamefont{Grosshans, Cerf,
  Wenger, Tualle-Brouri, and Grangier}}]{Grosshans2003a}
\bibinfo{author}{\bibfnamefont{F.}~\bibnamefont{Grosshans}},
  \bibinfo{author}{\bibfnamefont{N.~J.} \bibnamefont{Cerf}},
  \bibinfo{author}{\bibfnamefont{J.}~\bibnamefont{Wenger}},
  \bibinfo{author}{\bibfnamefont{R.}~\bibnamefont{Tualle-Brouri}},
  \bibnamefont{and} \bibinfo{author}{\bibfnamefont{P.}~\bibnamefont{Grangier}},
  \bibinfo{journal}{Quantum Inf. Comput.} \textbf{\bibinfo{volume}{3}},
  \bibinfo{pages}{535} (\bibinfo{year}{2003}).

\bibitem[{\citenamefont{Usenko and Filip}(2011)}]{Usenko2011}
\bibinfo{author}{\bibfnamefont{V.~C.} \bibnamefont{Usenko}} \bibnamefont{and}
  \bibinfo{author}{\bibfnamefont{R.}~\bibnamefont{Filip}},
  \bibinfo{journal}{New J. Phys.} \textbf{\bibinfo{volume}{13}},
  \bibinfo{pages}{113007} (\bibinfo{year}{2011}).

\bibitem[{\citenamefont{Derkach et~al.}(2017)\citenamefont{Derkach, Usenko, and
  Filip}}]{Derkach2017}
\bibinfo{author}{\bibfnamefont{I.}~\bibnamefont{Derkach}},
  \bibinfo{author}{\bibfnamefont{V.~C.} \bibnamefont{Usenko}},
  \bibnamefont{and} \bibinfo{author}{\bibfnamefont{R.}~\bibnamefont{Filip}},
  \bibinfo{journal}{Phys. Rev. A} \textbf{\bibinfo{volume}{96}},
  \bibinfo{pages}{062309} (\bibinfo{year}{2017}).

\bibitem[{\citenamefont{Lodewyck et~al.}(2007)\citenamefont{Lodewyck, Bloch,
  Garc{\'\i}a-Patr{\'o}n, Fossier, Karpov, Diamanti, Debuisschert, Cerf,
  Tualle-Brouri, McLaughlin et~al.}}]{Lodewyck2007}
\bibinfo{author}{\bibfnamefont{J.}~\bibnamefont{Lodewyck}},
  \bibnamefont{et~al.}, \bibinfo{journal}{Phys. Rev. A}
  \textbf{\bibinfo{volume}{76}}, \bibinfo{pages}{042305}
  (\bibinfo{year}{2007}).

\bibitem[{\citenamefont{Usenko and Filip}(2010)}]{Usenko2010a}
\bibinfo{author}{\bibfnamefont{V.~C.} \bibnamefont{Usenko}} \bibnamefont{and}
  \bibinfo{author}{\bibfnamefont{R.}~\bibnamefont{Filip}},
  \bibinfo{journal}{Phys. Rev. A} \textbf{\bibinfo{volume}{81}},
  \bibinfo{pages}{022318} (\bibinfo{year}{2010}).

\bibitem[{\citenamefont{Braunstein}(2005)}]{Braunstein2005a}
\bibinfo{author}{\bibfnamefont{S.~L.} \bibnamefont{Braunstein}},
  \bibinfo{journal}{Phys. Rev. A} \textbf{\bibinfo{volume}{71}},
  \bibinfo{pages}{055801} (\bibinfo{year}{2005}).

\bibitem[{\citenamefont{Leverrier et~al.}(2010)\citenamefont{Leverrier,
  Grosshans, and Grangier}}]{Leverrier2010}
\bibinfo{author}{\bibfnamefont{A.}~\bibnamefont{Leverrier}},
  \bibinfo{author}{\bibfnamefont{F.}~\bibnamefont{Grosshans}},
  \bibnamefont{and} \bibinfo{author}{\bibfnamefont{P.}~\bibnamefont{Grangier}},
  \bibinfo{journal}{Phys. Rev. A} \textbf{\bibinfo{volume}{81}},
  \bibinfo{pages}{062343} (\bibinfo{year}{2010}).

\bibitem[{\citenamefont{Ruppert et~al.}(2014)\citenamefont{Ruppert, Usenko, and
  Filip}}]{Ruppert2014}
\bibinfo{author}{\bibfnamefont{L.}~\bibnamefont{Ruppert}},
  \bibinfo{author}{\bibfnamefont{V.~C.} \bibnamefont{Usenko}},
  \bibnamefont{and} \bibinfo{author}{\bibfnamefont{R.}~\bibnamefont{Filip}},
  \bibinfo{journal}{Phys. Rev. A} \textbf{\bibinfo{volume}{90}},
  \bibinfo{pages}{062310} (\bibinfo{year}{2014}).

\bibitem[{\citenamefont{Madsen et~al.}(2012)\citenamefont{Madsen, Usenko,
  Lassen, Filip, and Andersen}}]{Madsen2012}
\bibinfo{author}{\bibfnamefont{L.~S.} \bibnamefont{Madsen}},
  \bibinfo{author}{\bibfnamefont{V.~C.} \bibnamefont{Usenko}},
  \bibinfo{author}{\bibfnamefont{M.}~\bibnamefont{Lassen}},
  \bibinfo{author}{\bibfnamefont{R.}~\bibnamefont{Filip}}, \bibnamefont{and}
  \bibinfo{author}{\bibfnamefont{U.~L.} \bibnamefont{Andersen}},
  \bibinfo{journal}{Nature Communications} \textbf{\bibinfo{volume}{3}},
  \bibinfo{pages}{1083} (\bibinfo{year}{2012}).

\bibitem[{\citenamefont{Weedbrook et~al.}(2004)\citenamefont{Weedbrook, Lance,
  Bowen, Symul, Ralph, and Lam}}]{Weedbrook2004}
\bibinfo{author}{\bibfnamefont{C.}~\bibnamefont{Weedbrook}},
	  \bibnamefont{et~al.}, 
  \bibinfo{journal}{Phys. Rev. Lett.} \textbf{\bibinfo{volume}{93}},
  \bibinfo{pages}{170504} (\bibinfo{year}{2004}).

\end{thebibliography}

\end{document}